\documentclass[useAMS,usenatbib]{mn2e}

\usepackage{graphicx}

\usepackage{natbib}

\title[A CH star in the globular cluster NGC 6426]{A CH star in the globular cluster NGC 6426}
\author[M. Sharina et al.]{M. Sharina$^{1}$\thanks{E-mail:
sme@sao.ru (MS)}, 
 B. Aringer$^{2}$, E. Davoust$^{3}$, A. Y. Kniazev$^{4,5,6}$, C. J. Donzelli$^{7,8}$\\
$^{1}$Special Astrophysical Observatory, Russian Academy of
Sciences, N. Arkhyz, KCh R, 369167, Russia\\
$^{2}$University of Vienna, Department of Astronomy, T\"urkenschanzstr. 17, A-1180 Wien, Austria \\
$^{3}$IRAP, Universit\'e de Toulouse, CNRS, 14 Avenue Edouard Belin, 31400 Toulouse, France \\
$^{4}$South African Astronomical Observatory, PO Box 9, 7935 Observatory, Cape Town, South Africa \\
$^{5}$Southern African Large Telescope Foundation, PO Box 9, 7935 Observatory, Cape Town, South Africa \\
$^{6}$Sternberg Astronomical Institute, Lomonosov Moscow State University, Moscow, Russia\\
$^{7}$Instituto de Investigaciones en Astronom\'ia Te\'orica y Experimental (IATE),\\ $ $ Observatorio Astron\'omico OAC, Laprida 854, X5000BGR, C\'ordoba, Argentina\\ 
$^{8}$Consejo Nacional de Investigaciones Cient\'ificas y T\'ecnicas (CONICET),\\ $ $ Avenida Rivadavia 1917, 
C1033AAJ, Buenos Aires, Argentina\\}

\begin{document}

\date{Accepted . Received ; in original form }


\maketitle

\label{firstpage}

\begin{abstract}
We report on the serendipitous discovery of a carbon star near the centre of the low-metallicity 
globular cluster NGC 6426.  We determined its membership and chemical properties using medium-resolution 
spectra. The radial velocity of -159 km/s makes it a member of the cluster.
We used photometric data from the literature and the COMARCS stellar atmospheric models to 
derive its luminosity, effective temperature, surface gravity, metallicity, and approximate 
C, N, and O abundance ratios. According to these properties, we suggest that this star is a genuine 
carbon rich low-metallicity AGB star.
\end{abstract}

\begin{keywords}
Stars: carbon - globular clusters: individual: NGC 6426
\end{keywords}

\section{Introduction}

The role of globular clusters (GCs) in the chemical evolution of galaxies
will be better understood by studying carbon stars (CSs) in the former. 
Only two types of CSs have abundances and kinematics typical for
the Galactic halo: the giant and dwarf CH stars \citep{mclure85, green96}. 
In the field most of these stars are binaries, where
the primary owes its peculiar nature to mass transfer from its white-dwarf companion
rather than to dredge-up from the interior \citep{mclure90}. 
There is no observational evidence for the exact nature of CH stars in GCs.
The chemical differences between CH stars that are intrinsic asymptotic giant branch (AGB)
stars and those in binary systems are detailed by \citet{abia03}.

Only three CH stars were found in GCs so far: two in $\Omega$ Cen,
RGO55 and GRO70 \citep{harding62, dickens72}, 
and a probable one in NGC 6402 \citep{cote97}.
The two CH stars in $\Omega$ Cen cannot be in the horizontal-branch phase
\citep{abia03}. 
Both GCs are of low-metallicity, and presumably belong to the Galactic halo.
Searches for CH stars in other clusters have been 
unsuccessful \citep{palmer80, palmer82}. 

We report on the serendipitous discovery of a CH star in a globular cluster,
and analyse its spectra to establish its properties and find clues
as to its exact spectral type.

\section{Observations and data reduction}

The CH star is located about $24.5\arcsec$ or $1.6 r_c$ south-east of the 
centre of NGC 6426, a low-metallicity GC in the Galactic halo.
Its coordinates, obtained with HST images and in the 2MASS all-sky point 
source catalogue\footnote{http://irsa.ipac.caltech.edu}, are: 
RA(J2000)$ = 17^h 44^m 55\fs 05$; Dec(J2000)$ = +03\degr10\farcm 09\farcs8$. 

The star was discovered during an observing run on 10 June 2010, 
at the 1.93m telescope of Observatoire de Haute-Provence (OHP) aimed at 
obtaining integrated spectra of Galactic globular clusters 
with the CARELEC spectrograph \citep{lemaitre90} and the grating 300 lines/mm.
Two long-slit ($5.5 \arcmin \times 2 \arcsec $) spectra
were obtained at the same position and PA=145.7$\degr$, both 
with 20-minute exposures and a seeing $\sim 2.5 \arcsec$. 
The corresponding dispersion and the spectral resolution were $\sim 1.78$\AA/pixels, 
and $\sim 5$\AA, and the spectral range was $\sim 3700 - 6800$\AA.
Helium and Neon lamps were exposed at the beginning and the end of the night 
for wavelength calibration. For relative flux calibration and radial velocity
calibration one flux standard HR8634 \citep{hamuy92, hamuy94}, two
Lick standard stars (HR5933 and HR7030) from the list of \citet{worthey94}, 
and an N-type CS (APM 2229+1902) from the list of \citet{totten98} 
were observed on the same night. 
The signal-to-noise per resolution element at 5000\AA\ is S/N$\sim$29. 

In order to reach a better S/N in the blue and to benefit from a higher spectral resolution, 
additional long-slit spectra were obtained with the Southern African Large Telescope (SALT) 
\citep{buckley06, odonoghue06} on 12 and 20 April 2012. 
We used the Robert Stobie Spectrograph \citep{burgh03} to make two 600 s. 
exposures with the grating GR900 and two 900 s. ones with GR2300. The slit width was 
1.25\arcsec\ and the position angle PA=146\degr. The final reciprocal dispersions are 1.78 and 
0.34 \AA\ pixel$^{-1}$ and the corresponding spectral resolution FWHM$=3$ and 1 \AA\ for the spectra 
taken with the two gratings, respectively.
The reduction of the SALT long-slit data was done in the way
described in \citet{kniazev08}.
Primary reduction of the data was done with the SALT science
pipeline \citep{crawford10}. 

The data reduction and analysis of the OHP observations were performed using 
MIDAS and IRAF.
The dispersion solution provides an accuracy of  $ \sim$0.08 \AA\ for the wavelength 
calibration.  
Possible systematics in the wavelength calibration due to instrumental flexure were
studied using the [OI]$\lambda5577$ night sky line in the dispersion-corrected spectra,
and by comparison of synthetic and observed spectra using the full spectrum fitting
methods and constructing a detailed line-spread function of the spectrograph (LSF) \citep[see e.g.][]{koleva09}.  

\begin{figure}
\hspace{1cm}
\centering
\includegraphics[width=\columnwidth,angle=-90,scale=0.73]{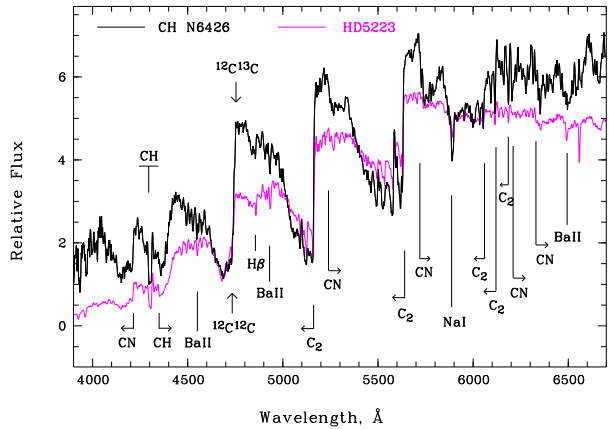}
\caption{Spectrum of the CH star.
The main spectral features are indicated. The spectrum of a classic CH star HD5223 from
Goswami (2005) is shown in gray for comparison.}
\label{fig_spectrum}%
\end{figure}

Finally the one-dimensional lower-resolution SALT spectrum was flux calibrated using the corresponding 
OHP spectrum and summed with it. The S/N in the resulting spectrum reaches $\sim$15 at 4300 \AA, 
$\sim$50 at 5000 \AA\, and $\sim$160 at 6300 \AA. The resolution of the spectrum obtained with the grating
GR2300 is FWHM$=1$\AA\ and the mean S/N at 4740 \AA\ is $\sim 40$.
The final medium-resolution spectrum of the CS is shown in Fig.\ref{fig_spectrum}.
The high-resolution spectrum was used separately to estimate $C ^{12}/C ^{13}$, 
as explained in Sect.~5.

\section{Radial velocity}

\begin{figure}
\centering
\includegraphics[width=\columnwidth,angle=-90,scale=0.73]{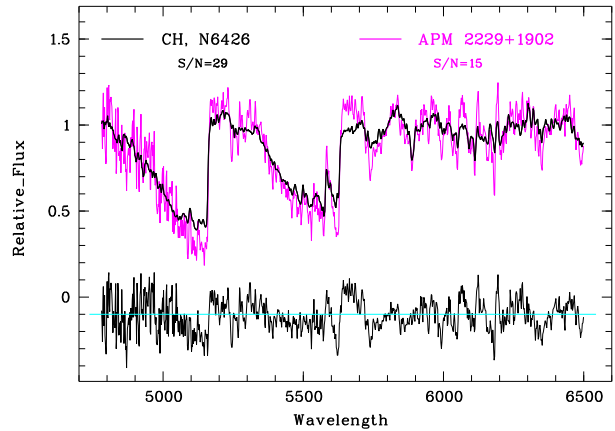}
\caption{Radial velocity determination for the CH star using the method {\sc pPXF}.}
\label{fig_comp}
\end{figure}

The heliocentric radial velocity of our star was 
derived by cross-correlation with the spectrum of the star APM 2229+1902 using 
the {\sc iraf} {\sc fxcor} package.
Before cross-correlation we subtracted the continuum using 
the {\sc iraf} task {\it continuum} and high-order polynomials. 
The cross-correlation shift (object-template) is 1.32 pixels. 
The radial velocity of our star is $V_h=-163 \pm 26$ km/s,
taking into account a relative systematic 
shift (object-template) derived from fitting of the night sky lines 
(-1.2 - (-0.7)$=$-0.5 pix), and
heliocentric corrections of -26 and -10 km/s, respectively,  for our star and the 
template APM 2229+1902 (which has a radial velocity $V_h=$-348 km/s).
We also used a more robust {\sc pPXF}
method \citep{cappellari04} to measure the radial velocity on the
same spectra. The result of the fitting is illustrated in Fig.\ref{fig_comp}.
Maximum penalized likelihood suppresses the effect of noise in the
solution, so the accuracy is higher. Applying the aforementioned corrections, 
the resulting heliocentric radial velocity is  $V_h=-159 \pm 5$ km/s. 
Both estimates are close to the radial velocity of NGC 6426 $V_h=-162 \pm 23$ \citep[][2010 edition]{harris96}.
The probability that the CH star belongs to the field is very low, because 
using the Besan\c con model \citep{robin03},
we find that the distribution of radial velocities of the 428 stars in the V magnitude range 10 to 18
located within 10\arcmin\ of the centre of the cluster 
is centered around a mean velocity of +13.22 km/s, with a dispersion of 53.52 km/s.

\section{Photometric information}

The photometric data for our CH star and the ones in NGC 6402 and $\Omega$ Cen,
and two classical Galactic CH stars from the atlas of CSs by \citet{barnbaum96}
are listed in Table~1. 
The successive lines are:
(1) apparent visual distance modulus from \citet[][2010 edition]{harris96}
(2) absolute magnitude in the V-band corrected for Galactic extinction, 
(3) Galactic extinction from \citet{schlegel98},
(4-9) broad-band optical and infrared colors corrected for Galactic extinction, 
(10) effective temperature.

To estimate the coordinates of the CH star in NGC 6402, we first obtained approximate coordinates  
on images from the CFHT archive\footnote{http://www1.cadc-ccda.hia-iha.nrc-cnrc.gc.ca},
and then identified it more accurately on HST images. These coordinates are: RA(J2000)$ = 17^h 37^m 36\fs94$;
Dec(J2000)$ = -03\degr14\farcm 5\farcs31$.
The colors and magnitudes for stars in NGC 6426 in the optical
bands were published by \citet{hatzi99} and \cite{dotter11}, 
using ground-based and HST images, respectively. 
Unfortunately, our star is saturated on the HST images.
In the color-magnitude diagram of \citet{hatzi99} it is the reddest star
in the cluster, slightly brighter than the stars at the tip of the red giant branch.
Photometric data in the B, V, and I bands for the CH stars in NGC 6402 and $\Omega$ Cen
were taken from \citet{cannon73}, \cite{cote97} and \cite{pancino07}. 
Unfortunately, the CH star in NGC 6402 is too close to another star of same magnitude
to obtain reliable near-infrared magnitudes from either 2MASS or 
Denis\footnote{http://cdsweb.u-strasbg.fr/denis.html}.
The data for HD5223 and V Ari were extracted from the SIMBAD astronomical database.

\begin{table}
\centering
\begin{minipage}{\columnwidth}
\caption{Optical and 2MASS photometric data for CH stars.
See text for details. All the colors were
corrected for Galactic extinction.}
\scriptsize
\begin{tabular}{lcccccl}
\hline
    &{\tiny N6426} &{\tiny N6402} &{\tiny RGO 55} &{\tiny RGO 70} &{\tiny  HD5223} &{\tiny  V Ari} \\
\hline
$(M-m)_0$ &16.58     &16.71     &13.97     &13.97     &  9.75     & 9.19     \\
$M_V$     &-2.58     &-1.86     &-2.39     &-2.36     & -1.43     & -1.14   \\
E(B-V)    &0.36      &0.60      &0.12      & 0.12     & 0.04      &  0.14    \\
B-V       &1.69      &1.90      &1.56      &1.68      & 1.43      & 2.14    \\
V-I       &1.59      &  -       &1.45      &1.31      &  -        &  -      \\
V-K       &3.52      &  -       & -        & -        & 2.82      & 3.73    \\
J-H       &0.69      &  -       &0.67      &0.73      &0.53       & 0.48    \\
H-K       &0.18      &  -       &0.23      &0.20      &0.17       & 0.26    \\
J-K       &0.87      &  -       &0.90      &0.93      &0.70       & 0.66    \\
$T_{eff}$, K &4100   &  -       &4468      &4287      &4360       & 3600    \\
\hline
\end{tabular}
\end{minipage}
\label{t1}
\end{table}

We calculated the near-infrared colors of the stars in the {\it SAAO} and {\it TCS} 
photometric systems using the 2MASS colors and the transformation equations 
from \citet{carpenter01} and \cite{ramirez04}.
With the colors thus derived, corrected for Galactic extinction, our star 
definitely falls in the box delimiting CH stars in the two-color diagnostic
diagram in the SAAO photometric system by \citet{totten00}.
Its effective temperature calculated using relations from \citet{alonso99}
and the $(J - K)_o$ color in the TCS photometric system is $T_{eff} = 4100 \pm 125$ K.
This corresponds to the class CH4 of \citet{keenan93}, which is equivalent to
spectral type K4 III for oxygen stars.
Similarly, the $T_{eff}$ of RGO 55 and RGO 70 are 4468K and 4287K, respectively.

\section{Spectral analysis}

\subsection{Classification}

The spectrum of the CH star was examined visually in terms of different
spectral characteristics \citep{goswami05} to avoid possible misclassification with 
C-N and R type CSs, which have spectra similar to those of CH stars. 
We compared the spectrum of the CH star with spectra at almost the same resolution for 
other such objects in the literature \citep[e.g][]{barnbaum96, goswami05}
and found an approximate similarity in the shape of the main spectral features with the spectra 
of two classical CH stars: HD5223 (type C-H3, C$_2$ index 4.5)
and HD13826 (V Ari) (type C-H3.5, C$_2$ index 5.5). HD 5223 is a CH giant 
(see Fig.~\ref{fig_spectrum})
with [Fe/H]$=-2.0$ dex, C/O$ = 3.0$, V Ari is a semi-regular pulsating star with 
[Fe/H]$=-2.4$, $log(g)=-0.2$, C/O$=2$, and $ ^{12}C / ^{13}C \sim 6 - 10$ 
\citep{aoki97, goswami05}.
The strength of {\it the G band of CH ($\sim$4300 \AA)} in the spectrum of our star,
{\it a main characteristic feature of CH-type CSs}, resembles that in the spectrum of HD 5223,
and the same applies to the
second branch near 4342 \AA. The line at 4226 \AA\ is very weak. It is blended by molecular bands. 
The lines of atomic hydrogen and BaII are seen distinctly, which is not the case in C-R stars.
The intensity of the CN band in our star is larger than in the case of HD 5223 and V Ari.  
The C$_2$ molecular bands near 4737, 5165, 5635, 6052 \AA\
are deeper than in HD 5223. This indicates a lower temperature for our star and/or
a higher C/O ratio \citep{barnbaum96, goswami05}.

CH stars are classified into two types which follow distinct evolutionary paths 
based on their  $ ^{12}C / ^{13}C$ ratios \citep[e.g][]{goswami05}.
Late-type CH stars with $^{12}C / ^{13}C \ge 100 $ are 
intrinsic asymptotic branch (AGB) stars that produce the s-process elements internally. 
Early-type CH stars with low values $^{12}C / ^{13}C \le 10 $ obtain the s-process elements
via binary mass transfer. This is why the $^{12}C / ^{13}C$ ratio is
an important probe of stellar evolution.
The band heads for $ ^{12}C ^{12}C$, $ ^{12}C ^{13}C$ and $ ^{13}C ^{13}C$ 
(at 4737\AA, 4744\AA\ and 4752\AA, respectively) are resolved even at medium resolution.  
In the following we will estimate $^{12}C / ^{13}C$ and other
parameters as accurately as possible using model atmospheres.

\subsection{Hydrostatic dust-free models}

\begin{figure}
\hspace{1cm}
\centering
\includegraphics[width=\columnwidth,angle=-90,scale=0.73]{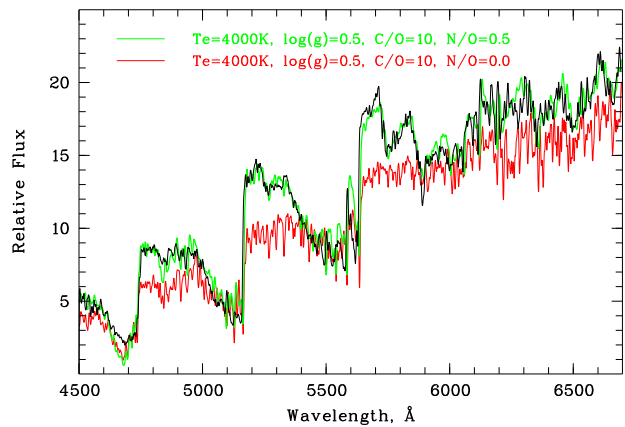}
\caption{Observed medium-resolution spectrum of the CH star in NGC 6426 (black) fitted with the model
one (green): $T_{eff}=4000$K, $log(g)=0.5$, C/O$=10$, N/O$=+0.5$. A model spectrum with lower
N abundance is shown in red.}
\label{spec_comp}%
\end{figure}

\begin{figure}
\hspace{1cm}
\centering
\includegraphics[width=\columnwidth,angle=-90,scale=0.7]{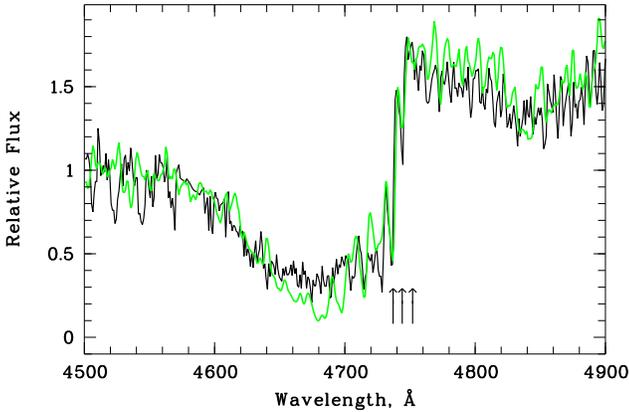}
\caption{SALT spectrum of the CH star in NGC 6426 taken at the resolution of FWHM$\sim 1$\AA.
The model with $T_{eff}=4000$K, $log(g)=0.5$, C/O$=10$, N/O$=+0.5$ is shown for comparison in grey. 
The band heads of $^{12}C ^{12}C$, $ ^{12}C ^{13}C$ and $ ^{13}C ^{13}C$ are indicated by arrows.}
\label{saltspectrum}%
\end{figure}

To estimate the effective temperature, surface gravity, C/O and $ ^{12}C / ^{13}C$ for the CH star
in NGC 6426 we used the hydrostatic dust-free models for carbon-rich giants of \citet{aringer09}
extended towards higher  $T_{eff}$, $log(g)$, and C/O to fit the observational data. 
We chose the following model parameters: stellar mass $1 M_{\sun}$, metallicity [Fe/H]$= -2.0$ dex,
and solar abundance for the other elements including  the isotopic ratio $ ^{12}C / ^{13}C=86.8$ 
\citep{scott06}. The metallicity was chosen equal to the  metallicity of NGC 6426 \citep[given by ][2010 edition]{harris96}.
The model stellar mass roughly corresponds to the mass of stars at the Main-Sequence turnover 
point of NGC 6426 \citep{hatzi99}. Unfortunately, 
it is not possible to derive more accurate values of the mass 
and chemical abundance of the star, given the resolution of our spectra.

The synthetic spectra computed with the COMA code based on COMARCS model atmospheres 
\citep[see][]{aringer09}  were degraded by convolution
with a Gaussian function to fit the resolution of our data. 
The resulting synthetic colors were compared with the observed ones, providing  
additional arguments for a proper evaluation of the stellar parameters. 

To obtain a good fit of the observational data with the 
model (both photometric and spectroscopic) 
we had to change not only the abundance
of Carbon, but also those of N and O, because only increasing C at a given  
$T_{eff}$, $log(g)$ did not allow us to fit the CN bands. 
A moderate increase of O with dredge up or mass transfer is expected for carbon stars
\citep[e.g][and references therein]{lugaro12}.
Finally, we derived the following model parameters: 
$T_{eff}=4000$K, $log(g)=0.5$, C/O$=10$, [N/Fe]$=+0.5$ dex, [O/Fe]$= +0.5$ dex
which corresponds to an absolute magnitude $M_V=-2.57$ and the following synthetic colors: 
$B-V=1.71$, $V-K=3.44$, $V-I=1.57$,
$J-H=0.67$, $H-K=0.24$, $J-K=0.91$.
One can see from Table~1 that the agreement between the observed and model spectra and 
between the spectroscopic and photometric $T_{eff}$ has been reached in a wide spectral range.

 Fig.~\ref{spec_comp} shows how this model fits our medium-resolution spectra.
Although our aim was not to estimate the abundances of different chemical elements on
these spectra,
the depth and shapes of the main molecular bands are fairly well adjusted. 
  We did not vary the isotopic ratio $ ^{12}C / ^{13}C$, 
since there is no significant difference between the shape and depth of the 
C$_2$ bands in the model and those in the spectrum
in the range 4737 - 4752 \AA . 
 Fig.~\ref{saltspectrum} shows the comparison between
the higher resolution spectrum of the CH star and the best fit model. 
The increased N abundance allows the strong CN features to be better matched. 
The derived luminosity, $T_{eff}$, $log(g)$ and the  high isotopic ratio $ ^{12}C / ^{13}C$ 
indicate that this is probably a {\it genuine} AGB star. The derived $T_{eff}$ is
normal for C-rich TP-AGB stars at a very low metallicity \citep{marigo08}.     

\section{Survival of binaries in globular clusters}

We now examine the alternative possibility that the CH star is a binary star
rather than a genuine AGB star.

The dense environment of GCs favors the formation of binary stars,
which in turn play an important role in their dynamical evolution, by delaying
the onset of core collapse \citep[e.g.][]{heggie06}. 
This dense environment may also lead to the dissolution of the wider binaries,
so that the actual number of binaries depends on the central stellar density
and core radius \citep{verbunt03}, as well as on their separation.

\begin{figure}
\centering
\includegraphics[width=\columnwidth]{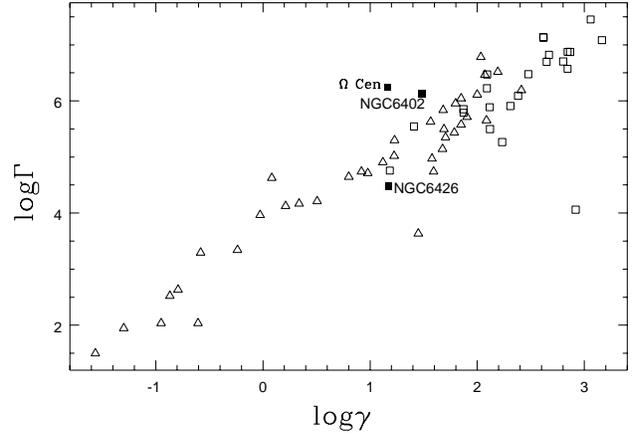}
\caption{Survival of a binary star. The total encounter rate $\Gamma$ is plotted vs the
encounter rate $\gamma$ for single binaries. Globular clusters with concentration parameters
c $<$ 1.60 are indicated by triangles, the others by squares.  The three globular clusters 
which harbour CH stars are indicated by full symbols. The units are arbitrary on both axes.}
\label{fig_gamma}
\end{figure}

In Fig.~\ref{fig_gamma} we show the total encounter rate 
$\Gamma \propto \rho_c^{1.5}ar_c^2$ 
versus the encounter rate $\gamma \propto \rho_c^{0.5}a/r_c$ for single binaries \citep{verbunt03}
for globular clusters in the halo (defined as having $[Fe/H] < - 0.80$ dex).
The parameters were calculated using the core radii $r_c$ and central 
luminosity densities $\rho_c$ from \citet[][2010 edition]{harris96} and the mass-to-light 
ratios from \citet{mclaughlin05}.
We assumed that the M/L of NGC 6426, not available in the catalogs, 
is $\sim$1.9, which is the most probable value for Galactic GCs at low metallicities. 
The uncertainty on this M/L is small (of the order of 10\%).
We also assumed that all binaries have the same semi-major axis $a$, which is a more
drastic simplification. 

The clusters are aligned along a relation of almost unit slope
between the two encounter rates.
The three GCs with known CH stars fall roughly in the middle of the relation.
If all four CH stars are binaries, this suggests
that no binaries can form when the total encounter rate is low, and
that they are more rapidly disrupted than replenished if the encounter rate for single
binaries is high.
The number of known CH stars in clusters 
is admittedly too small to allow for any statistical
inference on the probability of our CH star to be a binary.
The three GCs have similar $\gamma$, but 
the parameter $\Gamma$ is lower in NGC 6426 than in the two other 
clusters, thus making our star perhaps less likely to be a binary star. 
But, as pointed out by \citet{verbunt03}, $\Gamma$ depends 
on the mass segregation, on the fraction of binaries and on their period distribution.
Furthermore, both $\Gamma$ and $\gamma$ depend linearly on the semi-major axis,
for which we assumed a universal value, but which can vary between a few and at least 200 $R_\odot$,
so that presently a better way to determine the binary status of our star would be to monitor
its radial velocity.

\section{Conclusion}
We have discovered a Carbon star of CH type in NGC 6426 and derived its main physical and chemical parameters:
$M_V=-2.58$~mag, $T_{eff}=4000 - 4100$~K, 
$log(g) \sim 0.5$, [Fe/H]$\sim -2$~dex, C/O$\sim 10$, N/O$\sim+0.5$ dex, $ ^{12}C / ^{13}C \sim 87$.
The data and the estimated encounter rates in the parent GC indicate that the object is likely an {\it intrinsic} low-metallicity
carbon-rich AGB star, but
additional extensive high-resolution observational and theoretical studies are needed
to reveal the role of a possible presently invisible companion on its evolution.

\section*{Acknowledgments}
We thank Dr. Goswami for providing spectra of CH stars and 
Prof. Sarajedini for sending us a table with the photometry of NGC 6426,
and J.-P. Troncin, who helped in the OHP observations.
M. S. acknowledges partial support of grants GK. 14.740.11.0901, 
 and RFBG 11-02-00639-a.
BA acknowledges support from Austrian Science Fund
(FWF) Projects AP2300621 \& AP23586.
This research is based on observations obtained 
with the South African Large
Telescope, program 2011-3-RSA-003,
and it made use of the NASA/IPAC Infrared
Science Archive, which is operated by the Jet Propulsion
Laboratory, California Institute of Technology, under contract 
with the National Aeronautics and Space Administration.

\newpage
{\bf Note }:  After this paper was accepted for publication and posted on arXiv we were
informed that three more carbon star are known in the globular cluster $\Omega$ Cen 
(Jacco van Loon et al., MNRAS 382, 1353, 2007). 

\bsp

\label{lastpage}

\end{document}